\documentclass[showpacs,twocolumn,aps,amsmath,amssymb,prd]{revtex4}
\usepackage{epsfig}
\usepackage{wasysym}

\def\al{\alpha}
\def\be{\beta}
\def\ga{\gamma}
\def\de{\delta}

\def\et{\eta}

\def\ka{\kappa}
\def\la{\lambda}

\def\rh{\rho}

\def\ta{\tau}

\def\Ga{\Gamma}
\def\De{\Delta}

\def\mn{{\mu\nu}}

\def\fr#1#2{{{#1} \over {#2}}}

\def\frac#1#2{{\textstyle{{#1}\over {#2}}}}

\def\lsim{\mathrel{\rlap{\lower4pt\hbox{\hskip1pt$\sim$}}
    \raise1pt\hbox{$<$}}}
\def\gsim{\mathrel{\rlap{\lower4pt\hbox{\hskip1pt$\sim$}}
    \raise1pt\hbox{$>$}}}

\def\prt{\partial}

\def\etal{{\it et al.}}

\def\pt#1{\phantom{#1}}

\def\tt{$t^{\ka\la\mu\nu}$}
\def\uu{$u$}

\def\sss{s^{\mu\nu}}
\def\ttt{t^{\ka\la\mu\nu}}

\def\sb{\overline{s}}
\def\tb{\overline{t}}
\def\ub{\overline{u}}

\def\xx'{|\vec x -\vec x'|}

\def\pb{\overline {p}}

\newcommand{\beq}{\begin{equation}}
\newcommand{\eeq}{\end{equation}}
\newcommand{\bea}{\begin{eqnarray}}
\newcommand{\eea}{\end{eqnarray}}
\newcommand{\bit}{\begin{itemize}}
\newcommand{\eit}{\end{itemize}}
\newcommand{\rf}[1]{(\ref{#1})}

\begin{document}

\title{Time-delay and Doppler tests of the Lorentz symmetry
of gravity}

\author{Quentin G.\ Bailey}

\affiliation{Physics Department,
Embry-Riddle Aeronautical University,
3700 Willow Creek Road,
Prescott, AZ 86301, USA}

\date{\today}

\email{baileyq@erau.edu}

\begin{abstract}
Modifications to the classic time-delay effect and Doppler shift 
in General Relativity (GR) are studied in the context of the 
Lorentz-violating Standard-Model Extension (SME).
We derive the leading Lorentz-violating corrections to the 
time-delay and Doppler shift signals, 
for a light ray passing near a massive body. 
It is demonstrated that anisotropic coefficients 
for Lorentz violation control a time-dependent behavior 
of these signals that is qualitatively different from the 
conventional case in GR. 
Estimates of sensitivities to gravity-sector coefficients 
in the SME are given for current and future experiments, 
including the recent Cassini solar conjunction experiment.
\end{abstract} 


\maketitle

\section{Introduction}
\label{introduction}

At the present time, 
General Relativity (GR) remains the best known fundamental 
theory of gravity, 
describing all known classical gravitational phenomena.
Experiments testing this theory spanning $90$ years 
have failed to detect any convincing deviations.
Despite its continuing success, 
there remains widespread interest in pushing the limits 
of experimental tests of GR in order 
to find possible deviations.
This is primarily motivated by the consensus that there exists a 
unified fundamental theory that successfully meshes 
GR with the Standard Model of particle physics.
Such a theory may produce small deviations from GR that 
could manifest themselves in sensitive experiments.

One promising avenue of exploration involves 
searching for violations of the 
principle of local Lorentz symmetry \cite{cpt,reviews},
a foundation of GR.  
Candidate theories exist in which this symmetry principle
may be broken, 
at least at observable energy scales.
These scenarios include strings \cite{strings,ks1}, 
noncommutative field theories \cite{nc},
spacetime-varying fields \cite{spacetimevarying}, 
quantum gravity \cite{qg}, 
supersymmetric theories \cite{berger},
random-dynamics models \cite{fn},
multiverses \cite{bj}, 
and brane-world scenarios \cite{brane}.

A general theoretical framework for testing 
Lorentz symmetry in both gravitational 
and nongravitational scenarios has been developed
and is called the Standard-Model Extension (SME)
\cite{sme,akgrav}.
The SME is an effective field theory that 
incorporates the known physics of the 
Standard Model and GR,
while also including all possible Lorentz-violating terms
\cite{cptviolation}.
The Lorentz-violating terms are constructed from 
Standard Model and gravitational fields 
and coefficients for Lorentz violation,
which control the degree of the symmetry breaking.

One useful subset of the SME, 
called the minimal SME, 
contains the Lorentz-violating terms 
that dominate at low energies.
The matter sector of the minimal SME has been 
explored in experimental studies 
involving light \cite{light,light2,light3,light4,light5},
electrons \cite{electrons},
protons and neutrons \cite{protneut},
mesons \cite{mesons},
muons \cite{muons},
neutrinos \cite{neutrinos},
and the Higgs \cite{higgs}.
Some nonminimal SME terms, 
including Lorentz-violating operators of higher mass dimension, 
have already been explored in the photon sector in Refs.\ \cite{nonmin}.
In addition, 
because of the similarities of spacetime torsion to certain 
types of Lorentz violation,
experimental searches for SME coefficients have been used 
to place new torsion constraints \cite{torsion}.
A summary of the current experimental constraints on SME
coefficients can be found in Ref.\ \cite{tables}.

Studies of the curved spacetime generalization of the SME
have recently begun.
Within the setting of a general Riemann-Cartan spacetime, 
the dominant SME lagrangian terms in the matter and 
gravitational sector have been established \cite{akgrav}.
The matter sector of the SME couples to gravity
via the spin connection and vierbein.
In this scenario, 
some novel effects can occur that 
are controlled by certain matter sector coefficients 
which are unobservable in the flat spacetime limit
\cite{tkgrav}.
In the pure-gravity sector, 
key experimental signals in the Riemann spacetime 
limit have been established \cite{qbkgrav}. 
Experimental work constraining SME 
coefficients in the gravity sector
has already begun with atom-interferometric gravimeters
\cite{gravi}, 
lunar laser ranging \cite{llr}, 
Gravity Probe B \cite{gpb}, 
and short-range gravity tests \cite{srg}. 

Of the classic tests of GR,
the so-called fourth test, 
involving the measurement of the Shapiro time delay of light
passing near a massive body \cite{shapiro},
has recently gained attention. 
Improvements in two-way radio communication
with deep-space satellites, 
such as the Cassini probe, 
make possible a reduction in solar corona noise, 
yielding significant improvements in the accuracy of 
such tests \cite{bit}.
Further improvement in both time-delay and 
light-bending tests is also expected in the future
\cite{lator,odyssey,beacon,astrod,sims,st1,bepicolombo}.
It is therefore relevant to analyze 
in some detail the signals for Lorentz violation
in such experiments.
Some preliminary results describing the 
leading Lorentz-violating corrections to the 
Shapiro time-delay effect were 
obtained in Ref.\ \cite{qbkgrav} 
and were applied to the case of binary-pulsar timing experiments.
We seek here to elaborate on these results,
determine in addition 
the associated gravitational frequency shift signal,
and study potential signals in solar-system experiments. 

This paper is organized as follows.
In Sec.\ \ref{theory}, 
we discuss the theoretical foundations of this work.
Section \ref{basics} reviews key results in the gravitational
sector of the SME, 
including the post-newtonian metric.
We discuss light propagation in a background
spacetime in Sec.\ \ref{light propagation}, 
and apply the results to obtain the
time-delay formula in Sec.\ \ref{time delay}
and the frequency shift formula in Sec.\ 
\ref{frequency shift}.
In Sec.\ \ref{experiments}, 
we examine the results in the solar-system scenario.
Some preliminary discussion of the experimental
scenario in Sec.\ \ref{preliminaries} is
followed in Sec.\ \ref{time delay and Doppler signals}
by some exploration of the features 
of the Lorentz-violating signals in time-delay tests 
and Doppler tests. 
We discuss how analysis might proceed
and estimate sensitivities for existing
and future experiments in 
Sec.\ \ref{experimental analysis}. 
The main results of this work 
are summarized in Sec.\ \ref{summary}.
Throughout this work we adopt standard notation and conventions 
for the SME, 
as contained in Refs.\ \cite{sme,akgrav,qbkgrav}.
In particular, 
we work in natural units where $\hbar=c=1$
and with the metric signature $-+++$.

\section{Theory}
\label{theory}

\subsection{Basics}
\label{basics}

The SME with gravitational and nongravitational 
couplings was presented in Ref.\ \cite{akgrav}.
The general scenario is a Riemann-Cartan spacetime and includes 
couplings to curvature and torsion degrees of freedom. 
We focus here on the pure-gravity sector in the Riemann spacetime limit, 
within the minimal SME case.
The relevant action for this sector of the SME is written as
\bea
S &=& \fr {1}{16\pi G} \int d^4x \sqrt{-g}[(1-u)R + \sss R^T_\mn 
\nonumber\\
&& + \ttt C_{\ka\la\mu\nu}]+ S^\prime.
\label{act}
\eea
In this expression, 
$g$ is the determinant of the spacetime metric $g_\mn$,
$R$ is the Ricci scalar, 
$R^T_\mn$ is the trace-free Ricci tensor, 
$C_{\ka\la\mu\nu}$ is the Weyl conformal tensor, 
and $G$ is Newton's gravitational constant.  
The 20 coefficients for Lorentz violation 
\uu, 
$s^\mn$, 
and \tt control the 
leading Lorentz-violating gravitational couplings.
The additional piece of the action denoted $S'$
contains the matter sector and possible dynamical terms
governing the $20$ coefficients.

In the SME formalism, 
the action maintains general coordinate invariance, 
or \it observer \rm 
diffeomorphism symmetry,
as well as 
\it observer \rm local Lorentz symmetry.
However, 
because of the transformation properties
of the coefficients for Lorentz
violation, 
the SME action breaks both \it particle \rm 
local Lorentz symmetry and
\it particle \rm 
diffeomorphism symmetry \cite{akgrav,bkgrav}.
In the present context of the action in Eq.\ \rf{act}, 
the degree to which the particle symmetries 
are broken is controlled by the 
coefficients 
\uu, 
$s^\mn$, 
and \tt.

It has been demonstrated that explicit breaking 
of local Lorentz and diffeomorphism symmetry
generally conflicts with the Bianchi identities of Riemann geometry 
\cite{akgrav}.
In the action \rf{act} above,
explicit symmetry breaking would correspond to specifying 
\it a priori \rm 
the functional forms of the coefficients 
\uu, 
$s^\mn$, 
and \tt.
If the Lorentz-symmetry breaking is dynamical, 
however, 
the conflict with Riemann geometry is avoided \cite{akgrav}.
In the latter scenario the coefficients for Lorentz violation 
are dynamical fields and satisfy their own equations of motion.
This ensures that the Bianchi identities hold.

We consider here the case of spontaneous Lorentz violation. 
The dynamics governing the coefficients 
appearing in Eq.\ \rf{act} are contained in the $S'$ term. 
Through a dynamical process, 
the coefficient fields acquire vacuum expectation values 
that are denoted as
$\ub$, 
$\sb^\mn$,
and $\tb^{\ka\la\mu\nu}$.
For example, 
this may occur through the introduction of potential terms 
in $S'$ for 
\uu, 
$s^\mn$, 
and \tt,
whose minima are nonzero \cite{strings,ks1,akgrav,bkgrav,bkmass}.
This scenario has been treated for the action in Eq.\ \rf{act} 
in the linearized gravity limit, 
along with a broad study of signals for Lorentz violation 
in gravitational experiments, 
in Ref.\ \cite{qbkgrav}.
In particular, 
the post-newtonian metric was obtained, 
which comprises the starting point of this work.
Note that models of spontaneous Lorentz-symmetry breaking, 
capable of producing the effective coefficients for Lorentz
violation in \rf{act}, 
exist in the literature.
These include scalar \cite{scalar}, 
vector \cite{ks1,akgrav,bkgrav,bkmass,bb,ms}, 
and two-tensor models \cite{cardinal}. 

To study the propagation of light signals in a weak-field 
gravitational system, 
such as the solar system, 
the dominant $O(2)$ contributions to the post-newtonian metric are needed
\cite{PNorders} . 
The relevant terms in the metric for the pure-gravity sector 
of the minimal SME are controlled by the coefficients $\sb^\mn$.
They can be written in component form,
in an asymptotically inertial post-newtonian coordinate system
\cite{cartesian}, 
as
\bea
g_{00} &=& -1+ (2+3\sb^{00})U + \sb^{jk} U^{jk} + O(3),
\nonumber\\
g_{0j} &=& (a_1-2)\sb^{0j}U - a_1\sb^{0k}U^{jk} + O(3),
\nonumber\\
g_{jk} &=& \de^{jk} + [2+(1-2a_2)\sb^{00}]\de^{jk}U 
+ 2(a_2-1)\sb^{jk}U 
\nonumber\\
&&
\hskip-19pt
+[\sb^{lm}\de^{jk} 
-a_2 \sb^{jl}\de^{km}
-a_2 \sb^{kl}\de^{jm}
+ 2 a_2 \sb^{00}\de^{jl} \de^{km}]U^{lm}.\nonumber\\ 
\label{metric}
\eea
In the limit of vanishing $\sb^\mn$ coefficients,
the post-newtonian metric of GR is recovered.
The potentials appearing in this metric 
are given for an arbitrary mass density $\rh$ 
by
\bea
U &=& G \int \fr{\rh (\vec x', t)}{\xx'} d^3x',
\nonumber\\
U^{jk} &=& G \int \fr{\rh (\vec x', t) (x-x')^j (x-x')^k}
{\xx'^3} d^3x'.  
\label{monopole}
\eea
In Eqs.\ \rf{metric}, 
some coordinate gauge freedom remains in the
two quantities $a_1$ and $a_2$. 
For example, 
the standard harmonic gauge can be obtained
by setting $a_1=a_2=1$.
We leave these quantities unspecified to explicitly display the 
gauge-dependent nature of some of the results we derive in this work.
As discussed in detail elsewhere \cite{qbkgrav}, 
the relationship between this metric 
and the standard Parametrized Post-Newtonian (PPN) metric \cite{cmw}
is one of partial overlap in a special isotropic limit of the SME.

We will consider in this work the post-newtonian metric 
that is produced by a massive body at rest at the origin 
of the chosen coordinate system.
The dominant contributions to the potentials appearing in \rf{metric} 
are from the monopole terms. 
They depend on the coordinate position of the test body relative 
to the origin $r^j$. 
Thus we use
\bea
U &=& \fr {GM}{r}, 
\nonumber\\
U^{jk} &=& \fr{GM r^j r^k} {r^3}, 
\label{potentials}
\eea
where $M$ is the suitably defined mass 
of the central body.
In \rf{potentials}, 
we have neglected higher multipoles, 
which can play a role in systematics \cite{kop,st1}, 
and would be needed for a full treatment of the general
relativistic time-delay and Doppler shift signals.
For the present purposes, 
however, 
we need only the dominant contributions to these signals
that are controlled by the $\sb^{\mu\nu}$ coefficients.

If the mass of the central body is distributed significantly
outwards from its center,
then a substantial spherical moment of inertia
can arise, 
as happens with the Earth.
In this case, 
for a light signal grazing the surface of the central body, 
terms in the metric proportional to the 
moment of inertia $I$ of the body,
as well those that might be produced from a quadrupole moment, 
can give a significant contribution to resulting signals
controlled by the coefficients $\sb^\mn$ \cite{qbkgrav}.
For simplicity in this work, 
we neglect such cases and discard the metric terms
dependent on $I$.
This is not expected to produce a severe problem in 
typical solar-system experiments since it is known, 
in terms of the Sun's mass $M$
and radius $R_{\odot}$, 
that $I_{\odot}\approx 0.059 MR^2_{\odot}$ \cite{aa}. 

\subsection{Light propagation}
\label{light propagation}

To find both the time-delay signal and 
the Doppler shift signal we employ
standard methods and adopt the geometric optics
limit of electrodynamics in curved spacetime
\cite{mtw,bg1}.
We take the wave vector of a light ray,
tangent to the light path $x^\mu (\la)$,
to be
\beq
p^\mu = \fr{dx^\mu}{d\la},
\label{wavevec}
\eeq
where $\la$ is an affine parameter.
Since the light ray is a null geodesic, 
it obeys the geodesic equation and
the null vector condition given by
\bea
\fr {dp^\mu}{d\la} &=& -\Ga^\mu_{\pt{\mu}\al\be} p^\al p^\be,
\nonumber\\
p^\mu p^\nu g_\mn &=& 0.
\label{geod}
\eea
Note that under these assumptions we 
are neglecting Lorentz violation 
in the photon sector of the SME, 
which in any case is tightly constrained compared
to the gravitational sector 
\cite{light,light2,light3,light4,light5}.

We first consider a one-way light 
signal sent from an event $E$ to 
an event $P$, 
that passes a central body.
We will need to find the deviation of the light ray path 
in curved spacetime
from the straight line path in 
Minkowski spacetime between the two events.
The spatial endpoints of the path will be fixed 
at the two events $E$ and $P$, 
which amounts to solving \rf{geod} as 
a boundary-value problem rather than an 
initial-value problem.

To find the corrections due to curved spacetime 
we adopt a perturbative method using 
the linearized expansion for the metric and an expansion 
for the wave vector
\bea
g_\mn &=& \et_\mn+h_\mn,
\nonumber\\
p^\mu &=& \pb^\mu + \de p^\mu.
\label{wvexp}
\eea
Here $h_\mn$ are the metric fluctuations, 
representing the deviation of $g_\mn$ from 
the flat spacetime metric $\et_\mn$.
The first term in the second equation 
is the zeroth-order wave vector
that is constant and satisfies the 
condition $\et_\mn \pb^\mu \pb^\nu=0$.
The second term $\de p^\mu$ is the correction 
to the wave vector due to curved spacetime.
Applying the null condition for the full wave vector $p^\mu$
to leading order in the metric perturbation $h_\mn$
yields a constraint on $\de p^\mu$:
\beq
2\pb^\mu \de p^\nu \et_\mn \approx -h_\mn \pb^\mu \pb^\nu.
\label{constr1}
\eeq

\begin{figure}[h]
\begin{center}
\epsfig{figure=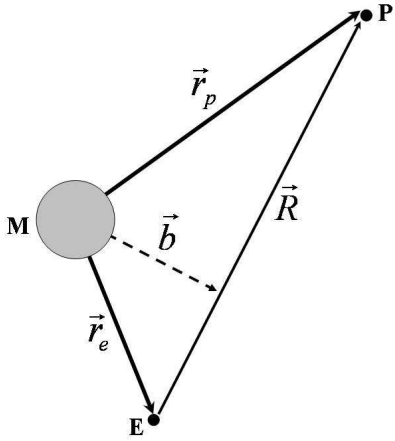,width=0.98\hsize}
\caption{\label{fig1}
Diagram illustrating the meaning of
the position vectors $\vec r_p$, $\vec r_e$, 
$\vec R$, 
and the impact parameter $\vec b$, 
for the zeroth-order light trajectory
passing a massive body from events E
to P.}
\end{center}
\end{figure}

We shall denote the coordinates
of the endpoint events $E$ and $P$ as $(t_e, r_e^j)$ 
and $(t_p, r_p^j)$, 
respectively.
Generally in what follows, 
quantities referred to each of the events
are denoted with subscripts $e$ and $p$.
The zeroth-order spatial trajectory for the light 
ray will be a straight line in the direction 
$\vec R = \vec r_p - \vec r_e$.
This implies that the zeroth-order wave vector, 
tangent to this straight line,
has components $\pb^0 = 1$ and $\pb^j=\hat R^j$, 
where $\hat R=\vec R/R$ and $R=|\vec R|$.
The zeroth-order spatial trajectory can be written as 
\beq
x_0^j (\la)= \hat R^j \la + b^j,
\label{traj}
\eeq
where $b^j$ is the impact parameter vector.
It can be written as
\beq
b^j = r_p^j - \hat R^j \vec r_p \cdot \hat R. 
\label{impact}
\eeq
Furthermore, 
to be consistent with the boundary conditions,
the parameter $\la$ is taken to vary from 
$-l_e=\vec r_e \cdot \hat R$
to $l_p=\vec r_p \cdot \hat R$, 
from which it follows that $l_e+l_p=R$.
The various quantities that we use to describe 
the zeroth-order trajectory of a light ray passing 
a central body are depicted in Fig.\ \ref{fig1}.

The parametrization and definitions above
have an immediate consequence on $\de p^j$.
Integration of the spatial components 
of the definition \rf{wavevec} over the light path, 
followed by use of the second equation in 
\rf{wvexp} yields
\beq
\int_{-l_e}^{l_p} \de p^j d\la=0.
\label{depj2}
\eeq
This result is just a reflection 
of the fact that the spatial endpoints
of the trajectory are fixed. 
Equation \rf{depj2} will be useful in deriving 
the light travel time formula and the Doppler shift formula, 
as we show below.

The corresponding integral involving the time component
$\de p^0$ does not vanish, 
however,
on account of its being fixed by Eq.\ \rf{constr1}.
In fact, 
it can be used to derive the light travel time.
Integrating the time component of the 
definition \rf{wavevec} from $E$ to $P$, 
and making use of \rf{constr1} and \rf{depj2}, 
we obtain
\bea
t_p - t_e = R + \fr 12 \int_{-l_e}^{l_p} h_\mn \pb^\mu \pb^\nu d\la.
\label{td1}
\eea
This equation forms the starting point for 
the derivation of the time-delay formula 
in Sec.\ \ref{time delay}.

We now consider the shift in the frequency of light 
measured by two observers at the two events $E$ and $P$.
The ratio of the frequencies $\nu$ measured at
the two events can be obtained from the standard formula
\beq
\fr {\nu_P}{\nu_E} = \fr {(U_p^\mu p_\mu)_P}{(U_e^\mu p_\mu)_E},
\label{fr1}
\eeq
where $U_p^\mu$ and $U_e^\mu$ are the four-velocities
of two distinct observers present at events $P$ and $E$, 
respectively. 
Note that the quantities in the numerator and denominator are
to be evaluated at the two events $P$ and $E$, 
as indicated.

To obtain an explicit expression for the 
frequency shift for the one-way trip past a massive body, 
it will be convenient to work with 
the covariant components of the wave vector 
$p_\mu = g_\mn p^\nu$.
Expanding Eq.\ \rf{fr1} into space and time
components yields
\beq
\fr {\nu_P}{\nu_E} = \left(\fr {dt}{d\ta_p}\right) 
\left(\fr {d\ta_e}{dt}\right) 
\left( \fr {p_0+w^j p_j}{p_0+v^j p_j}\right), 
\label{fr2}
\eeq
where $v^j$ and $w^j$ are the coordinate velocities
of the two observers at events $E$ and $P$, 
and $\ta_e$ and $\ta_p$ are their proper times,
respectively.

One convenient consequence of using the covariant
wave vector is that, 
for a spacetime metric
with no explicit time dependence, 
the component $p_0$ will be constant along
the path of light $x^\mu (\la)$.
Furthermore, 
this condition will be 
approximately true for the post-newtonian
metric \rf{metric} from a massive body 
approximately at rest at the origin of the 
chosen post-newtonian coordinate system. 
Therefore we take
\beq
\fr {d p_0}{d\la} \approx 0.
\label{p0}
\eeq
It will be important, 
however, 
to determine what the constant $p_0$ is, 
in order to obtain the correct
frequency shift result.

To determine the covariant components $p_0$
and $p_j$ of the wave vector in Eq.\ \rf{fr2} we first
expand in the manner of \rf{wvexp}:
\beq
p_\mu = \pb_\mu + \de p_\mu,
\label{wvexp2}
\eeq
where $\pb_\mu=\et_\mn \pb^\nu$. 
Using the null constraint \rf{geod}, 
Eq.\ \rf{constr1},
and the properties of $\pb^\mu$,
we can establish that
\bea
\de p_0 &=& -\hat R^j \de p^j + h_{0\mu} \pb^\mu 
- \fr 12 h_\mn \pb^\mu \pb^\nu,
\nonumber\\
\de p_j &=& \de p^j + h_{j\mu}\pb^\mu.
\label{covdeps}
\eea 

If we integrate the constant $p_0=-1+\de p_0$ 
over the light path, 
use the first equation in \rf{covdeps}, 
and Eq.\ \rf{depj2},
we can establish that
\beq
\de p_0 = \fr {1}{R} \int_{-l_e}^{l_p} 
\left(h_{0\mu} \pb^\mu - \fr 12 h_\mn \pb^\mu \pb^\nu \right)d\la.
\label{dep0}
\eeq
Furthermore, 
if we insert the expansion \rf{wvexp2}
into the geodesic equation \rf{geod}, 
and integrate over the path we find
\beq
\de p_j (P) - \de p_j (E) = 
\fr 12 \int_{-l_e}^{l_p} \prt_j h_\mn \pb^\mu \pb^\nu d\la.
\label{depj3}
\eeq

To find the value of $\de p_j$ at the endpoints, 
which is needed to evaluate the frequency shift \rf{fr2},
we start with the expression \rf{depj2}.
A suitable integration by parts, 
followed by the use of \rf{depj3}, 
yields the values of $\de p_j$ at the two events
$P$ and $E$ in terms of integrals of metric components:
\bea
\de p_j (P) &=& \fr {1}{2R} \int_{-l_e}^{l_p} 
[(\la + l_e)\prt_j h_\mn \pb^\mu \pb^\nu 
+ 2 h_{j\mu}\pb^\mu)] d\la,
\nonumber\\
\de p_j (E) &=& \fr {1}{2R} \int_{-l_e}^{l_p} 
[(\la - l_p)\prt_j h_\mn \pb^\mu \pb^\nu 
+ 2 h_{j\mu}\pb^\mu)] d\la.\nonumber\\
\label{depj4}
\eea
The expressions \rf{depj4} and \rf{dep0} form the starting
point of the derivation of the Doppler shift
in Sec.\ \ref{frequency shift}.
Note that, 
although we will focus 
in the next sections on the metric from 
the gravity sector of the minimal SME,
the results \rf{td1}, 
\rf{fr2}, 
\rf{dep0}, 
and \rf{depj4} could be applied to 
alternative theories of gravity in the linearized limit, 
with an approximately time-independent metric.
In particular, 
though it lies beyond the scope of
the present work, 
it would be of interest to investigate
effects outside of the gravity sector of the minimal SME, 
such as matter-gravity couplings \cite{tkgrav}.
Finally, 
we note in passing that our results in Eqs.\ \rf{depj4} 
are consistent with Ref.\ \cite{bg1}.

\subsection{Time-delay formula}
\label{time delay}

To establish the one-way light travel time, 
which contains a time-delay term due to curved spacetime, 
we must evaluate the integral in Eq.\ \rf{td1}.
The projection of the metric along $\pb^\mu$ that appears
in the integrand can be written as
\beq
h_\mn \pb^\mu \pb^\nu  = \De U + \De^{jk} U^{jk}.
\label{hpp}
\eeq
The quantities $\De$ and $\De^{jk}$ are given by
\bea
\De &=& 4 + \sb^{00} (4-2 a_2) + (2 a_1 -4)\sb^{0j} \hat R^j
\nonumber\\
&&
+2 (a_2-1) \sb^{jk} \hat R^j \hat R^k,
\nonumber\\
\De^{jk} &=& 2 \sb^{jk} - a_1 \sb^{0j}\hat R^k- a_1 \sb^{0k}\hat R^j
-a_2 \sb^{jl}\hat R^l \hat R^k
\nonumber\\
&&
-a_2 \sb^{kl}\hat R^l \hat R^j
+2 a_2 \hat R^j \hat R^k \sb^{00}.
\label{deltas}
\eea
With these definitions the light travel time takes the form
\beq
t_p - t_e  = R + \fr 12 \De \int^{l_p}_{-l_e} U d\la
+ \fr 12 \De^{jk} \int^{l_p}_{-l_e} U^{jk} d\la.
\label{td3}
\eeq

Using the monopole expressions in Eq.\ \rf{monopole}, 
and evaluating the potentials with the zeroth-order
spatial trajectory \rf{traj}, 
these integrals can be evaluated by standard methods.
The resulting expression for the one-way light travel time, 
to post-newtonian order $O(2)$,
is given by
\bea
t_p-t_e &=& R +2GM(1+\sb^{00}-\sb^{0j} \hat R^j) 
\ln \left[\fr {r_e + r_p + R}{r_e +r_p -R}\right] 
\nonumber\\
&& 
+GM[-a_2 \sb^{00}+ a_1 \sb^{0j} \hat R^j+\sb^{jk} \hat b^j \hat b^k
\nonumber\\
&&
\pt{M}
+(a_2-1)\sb^{jk}\hat R^j \hat R^k]
\left(\fr {l_e}{r_e} +\fr {l_p}{r_p}\right),
\nonumber\\
&&
+GM[a_1 \sb^{0j} b^j +(a_2-2)\sb^{jk} \hat R^j b^k] 
\fr {(r_e - r_p)}{r_e r_p}
\nonumber\\ 
&& 
+...,
\label{td4}
\eea
where the ellipses represent higher order 
post-newtonian corrections.
Neglecting these term suffices to establish the leading effects from 
Lorentz violation for experiments.
Note that this one-way result matches that obtained in 
Ref.\ \cite{qbkgrav} in the appropriate limit.
Also, 
in the isotropic limit of the SME,
and for the appropriate coordinate choice, 
the result \rf{td4} matches the standard PPN result \cite{mtw,cmw}.

In many practical cases, 
the light signal is reflected from a planet or spacecraft.
Using \rf{td4} we can establish the round-trip light travel time.
This involves adding the light travel time for a signal 
transmitted by an observer at event $P$ that 
travels to the other observer arriving at an event denoted $E'$.
The light travel time for the return trip 
can be obtained from \rf{td4} with the substitutions 
\bea
\vec r_e &\rightarrow& \vec r_p,
\nonumber\\
\vec r_p &\rightarrow& \vec r_e\,',
\label{subs}
\eea
where $\vec r_e\,'$ is the position of event $E'$.
Note that the quantities $\vec R$ and  
$\vec b$ will change for the return trip accordingly.
We assume that the observer at event $E$, 
later receiving the returned signal at event $E'$,
is traveling at small velocities
compared to $1$.
Thus it suffices to approximate the motion during the light
transit as rectilinear.
The small velocities are in any case 
implied by the post-newtonian expansion adopted here.
If we account for this motion during the light transit, 
but we neglect terms of order $GMv$, 
the order $GM$ portion of the light travel time is
equal to its value for the outgoing trip, 
except for sign changes in the $\sb^{0j}$ terms.
Thus we obtain for the round-trip light travel time  
\bea
\De t &\approx& \fr {2R(1-\hat R \cdot \vec v)}{1-v^2} 
+4GM(1+\sb^{00}) 
\ln \left[\fr {r_e + r_p + R}{r_e +r_p -R}\right]  
\nonumber\\
&& 
+2GM[-a_2 \sb^{00}+ \sb^{jk} \hat b^j \hat b^k
\nonumber\\
&&
\pt{2GM}
+(a_2-1)\sb^{jk}\hat R^j \hat R^k]
\left(\fr {l_e}{r_e} +\fr {l_p}{r_p}\right)
\nonumber\\
&&
+2GM (a_2-2)\sb^{jk} \hat R^j b^k 
\left(\fr {r_e - r_p}{r_e r_p}\right). 
\label{td5}
\eea
Note that the terms with the $\sb^{0j}$ coefficients 
canceled when adding the outgoing and return trip contributions. 
This is due to their oddness under parity.

Neglecting terms of order $GMv$,
the measured elapsed proper time $\De \ta_e$ at the receiver 
is related to the above result
by the factor $d\ta_e/dt$,
which is to be evaluated along the worldline of the receiver.
In principle, 
this factor contains contributions from the 
spacetime metric near the observer present at event $E$
and is related to the classic gravitational redshift as 
discussed in the next subsection.
For our analysis in this work, 
we focus on effects from a single body stemming
from the $O(GM)$ terms in the expression above, 
though the results could be generalized
to $N$ bodies.

There are two key time scales which could be used to 
distinguish the large special-relativistic effects contained in 
the first term in \rf{td5} from the smaller
terms of order $GM$ \cite{bg1}.
The time scale over which significant
changes occur with the first term is essentially
an orbital time scale $\overline{r}/\overline{v}$, 
where $\overline{r}$ and $\overline{v}$ are typical
orbital distances and velocities, 
respectively, 
comparable to $R$ and $v$ defined in Sec.\ \ref{light propagation}.
The conjunction time scale $b/\overline{v}$ is
approximately the time scale over which the $O(GM)$
terms in \rf{td5} vary significantly.
For typical experiments this is on the order of days.

The dominant contribution from the $O(GM)$ terms comes 
from the logarithmic term in \rf{td5}.
Note that the only coefficient for Lorentz violation
appearing in front of the logarithmic term is the 
rotational scalar $\sb^{00}$, 
which points to the possibility of its being measured
at the same level as the PPN parameter $\ga$.
Anisotropic coefficients control many of the
remaining terms.
As we show for specific experiments in Sec.\ \ref{experiments}, 
the typical size of the remaining terms are 
somewhat suppressed relative to the logarithmic term.

It is important to note that, 
in principle, 
the special-relativistic terms in \rf{td5} also 
receive corrections due to the gravity-sector 
coefficients $\sb^\mn$.
These corrections would arise
through modifications to the orbital dynamics of the transmitting and
reflecting bodies (e.g., Earth and spacecraft or planet).
For the purposes of detailed modeling, 
these effects could be included, 
for example, 
by modeling the orbits as oscillating ellipses.
Secular changes in the orbital elements due to 
the coefficients $\sb^\mn$ could be included 
using the results from Ref.\ \cite{qbkgrav}.
In any case, 
such orbital corrections are expected to be relevant over
the orbital time scale $\overline{r}/\overline{v}$.

\subsection{Frequency shift}
\label{frequency shift}

In GR, 
in addition to the bending of light 
and the time-delay effect, 
the frequency of light also changes after having passed
near a massive body \cite{shapiro66}. 
This effect, 
closely related to the time-delay effect, 
is distinct from the classic gravitational 
redshift and vanishes for stationary observers.
In this section, 
we evaluate the one-way frequency shift, 
using the results of Sec.\ \ref{light propagation},
and also determine the fractional
frequency shift for a two-way reflected signal.

We begin with Eq.\ \rf{fr2}, 
expanded to leading order in the wave vector
shift $\de p_\mu$:
\beq
\fr {\nu_P}{\nu_E} = \fr {dt}{d\ta_p} 
\fr {d\ta_e}{dt} 
\left( \fr {1-\de p_0-w^j \hat R^j - w^j \de p_j (P)}
{1-\de p_0-v^j \hat R^j-v^j\de p_j (E)}\right). 
\label{fr3}
\eeq
The result is expanded to all orders in velocity
for the special-relativistic terms, 
but to $O(3)$ in the terms depending 
on the metric fluctuations $h_\mn$.  
With some manipulation we obtain
\beq
\fr {\nu_P}{\nu_E} = \sqrt{\fr {1-\vec v^2}{1-\vec w^2}}
\fr {1-w^j \hat R^j}{1-v^j \hat R^j}
\left( 1+\left(\fr {\nu_P}{\nu_E}\right)_g
\right), 
\label{fr4}
\eeq
where the term arising from the effects of gravity
via the metric fluctuations is labeled $g$ and is given by 
\beq
\left(\fr {\nu_P}{\nu_E}\right)_g 
= \left(\fr {\nu_P}{\nu_E}\right)_{RS} 
+\left(\fr {\nu_P}{\nu_E}\right)_D.
\label{fr4p}
\eeq

The term labeled $RS$ on the right-hand side of \rf{fr4p} 
is the gravitational redshift.
This term arises from the spacetime metric being evaluated 
at the endpoints of the light trajectory, 
namely events $E$ and $P$.
It is given by
\bea
\left(\fr {\nu_P}{\nu_E}\right)_{RS} &=&
\left(1+\fr {3}{2} \sb^{00}\right) GM \fr {r_e-r_p}{r_e r_p}
\nonumber\\
&&
+\fr {1}{2}\sb^{jk} GM \left(\fr {r_p^j r_p^k}{r_p^3}
-\fr {r_e^j r_e^k}{r_e^3}\right)
+...,
\label{rs}
\eea
where the ellipses represents higher post-newtonian corrections.
Equation \rf{rs} includes leading Lorentz-violating
corrections to the standard gravitational redshift of GR, 
which is recovered in the limit $\sb^\mn=0$.
The result \rf{rs} would be of interest to investigate
for gravitational redshift experiments, 
such as those incorporating sensitive
atomic clocks on Earth or aboard orbiting satellites 
\cite{redshift,cmw,gps,aces}.
Our main focus in this work, 
however, 
will be on the time-delay effect and its associated contribution to 
the frequency shift derived below.

The term labeled $D$ in Eq.\ \rf{fr4p} is the 
gravitational frequency shift of light due to 
the wave vector corrections $\de p^\mu$, 
which reads
\begin{widetext}
\bea
\left(\fr {\nu_P}{\nu_E}\right)_D &=& 
\de p_0 (v-w)^j \hat R^j
-w^j \de p_j (P) + v^j \de p_j (E)
\nonumber\\
&=& \fr {1}{2R} \int_{-l_e}^{l_p} 
\big( [\la (v-w)^j -l_e w^j-l_p v^j)] \prt_j h_\mn \pb^\mu \pb^\nu
+ (v-w)^j \hat R^j ( 2 h_{0\mu} \pb^\mu - h_\mn \pb^\mu \pb^\nu)
+ (v-w)^j h_{j\mu} \pb^\mu \big) d\la.\nonumber\\
\label{fr5}
\eea
\end{widetext}
This term represents a gravitational correction
to the usual Doppler shift of Special Relativity.
The integrals in \rf{fr5} can be evaluated by inserting
the post-newtonian metric \rf{metric} and using the 
zeroth-order spatial trajectory of the light ray, 
in a manner similar to Sec.\ \ref{time delay}.
The result is significantly more cumbersome than \rf{td4}, 
and so we adopt an approximation that is suitable
for capturing the dominant terms that are proportional to 
the coefficients for Lorentz violation $\sb^\mn$.
After evaluating the integrals in \rf{fr5}, 
the results can be grouped according to 
powers of $(GM\overline{v}/b)(b/\overline{r})^n$. 
We will focus here on near-conjunction
time scales where the dominant terms in 
\rf{fr5} are of order $GM\overline{v}/b$ and higher
order terms will be suppressed by powers
of the small factor $b/\overline{r}$.

Keeping only the order $GM\overline{v}/b$ terms, 
we obtain for the frequency shift contribution 
\rf{fr5},
\beq
\left(\fr {\nu_P}{\nu_E}\right)_D \approx 
\fr{4GM}{b}
[ (1+\sb^{00}-\sb^{0j}\hat R^j +\sb^{jk}\hat b^j \hat b^k)\dot{b}
-\sb^{jk}\hat b^j \dot{b}^k].
\label{fr6}
\eeq
In this expression the dot denotes a time derivative
with respect to the post-newtonian coordinate time $t$.
Note that the arbitrary quantities $a_1$ and $a_2$
keeping track of the coordinate gauge freedom have
vanished in this result, 
indicating the coordinate invariance of \rf{fr6}.
The result \rf{fr6} can also be verified by taking 
the coordinate time derivative of \rf{td4} and using
a known relationship between the frequency shift
and light travel time \cite{shapiro66}.

The result \rf{fr6} can be contrasted with 
the contributions to the frequency shift contained in 
the remaining terms in Eq.\ \rf{fr4}
and also \rf{rs}.
In the same manner as the special-relativistic terms in 
the light travel time expression \rf{td5}, 
the velocity contributions in \rf{fr4} and 
the gravitational terms in \rf{rs}
will vary over the orbital time scale 
$\overline{r}/\overline{v}$ in typical experiments.
In contrast, 
the signal in \rf{fr6} will vary most significantly when the
light ray passes near the central body ($b<<\overline{r}$), 
when the observers and the central body are
in conjunction.

We now calculate the fractional frequency shift 
of a light signal reflected from the planet or spacecraft.
Thus we seek
\beq
\fr{\delta \nu}{\nu}=\fr{\nu'-\nu}{\nu},
\label{freq}
\eeq
where $\nu$ is the transmitted frequency 
and $\nu'$ is the returned frequency.
In a manner similar to what was done for the 
round-trip light travel time in Sec.\ \ref{time delay},
we can obtain the frequency shift for the return
signal with suitable substitutions in the one-way result
\rf{fr6}.
Adding the return signal to the outgoing one,
we find that the gravitational portion 
of the leading fractional frequency shift 
from the round-trip signal is given by
\beq
\left(\fr{\delta \nu}{\nu}\right)_g = 
\fr{8GM}{b}
[ (1+\sb^{00}+\sb^{jk}\hat b^j \hat b^k)\dot{b}
-\sb^{jk} \hat b^j \dot{b}^k ]+...,\\
\label{freq2}
\eeq
where the ellipses include terms of order 
$(GM\overline{v}/b)(b/\overline{r})$ and 
higher order post-newtonian corrections.
Note that the $\sb^{0j}$ terms have vanished due to
their oddness under Parity,
just as they did for the time-delay formula.
Also, 
the time dependence of \rf{freq2} is controlled by the 
behavior of the impact parameter vector $\vec b$
and its time derivative $\vec {\dot b}$.

\section{Experiments}
\label{experiments}

In this section, 
the experimental implications of the results derived
in Secs.\ \ref{time delay} and \ref{frequency shift} are 
examined in the context of key solar system experiments.
We point out the basic features of the 
Lorentz-violating signals and contrast them 
with the GR case.
Also, 
we describe how experiments could be used to 
probe various combinations 
of coefficients for Lorentz violation
and estimate the level of sensitivity for each test.

\subsection{Preliminaries}
\label{preliminaries}

We work in a post-newtonian coordinate system that
asymptotically coincides with the Sun-centered 
celestial-equatorial coordinate system adopted 
in most SME studies \cite{light4}.
Space and time coordinates in this system 
are denoted with capital letters $(T,X^J)$
\cite{scf}.
This approximation to an 
asymptotically inertial frame suffices 
for many SME experimental studies.
Note that the Sun's center is in orbit
around the barycenter of the solar system 
with a mean velocity about $1000$ times smaller
than the Earth's orbital velocity.
Standard practice in solar system experiments
is to adopt the Barycentric Celestial Reference System.
For our purposes here in identifying
the leading Lorentz-violating effects, 
it suffices to proceed in the Sun-centered frame
and neglect the Sun's motion. 
However, 
in establishing beyond leading order corrections
to the light travel time and Doppler observables 
in GR, 
the Sun's velocity can play a role \cite{sunvel,b07}.
 
To study the basic features of our results
we focus on the solar conjunction time scale 
where the signals for Lorentz violation 
are near their maximum.
In this scenario, 
where the light signal passes close to the Sun, 
we can assume approximately
rectilinear motion for the Earth observer and
the planet or spacecraft.
The main changing variable in this
case is the impact parameter vector \cite{bg1,bg2}.
Assuming rectilinear motion, 
we expand the impact parameter vector around
its minimum value $\vec b_0$ as
\beq
\vec b = \vec b_0 + \dot {\vec b}_0 T,
\label{impact2}
\eeq
where $\dot {\vec b}_0$ is the time derivative
of the impact parameter vector evaluated
at the conjunction time $T=0$. 
Note that we also have 
$\vec b_0 \cdot \dot {\vec b}_0=0$.

In many cases of interest, 
the time derivative of the impact parameter vector 
near the conjunction time can be written approximately as
\beq
\dot {\vec b}_0 \approx \fr {l_p \vec v + l_e \vec w}{R},
\label{bdot0}
\eeq
where $\vec v$ is the Earth receiver's velocity 
and $\vec w$ is the velocity of the spacecraft
or planet.
All quantities on the right-hand side of Eq.\ \rf{bdot0}
can be determined from their definitions in Sec.\ 
\ref{light propagation} and are evaluated at $T=0$.
Note that if the planet or spacecraft is many times
further from the Sun than the Earth,
so that $R>>l_e$ and $l_p \sim  R$, 
the primary contribution to \rf{bdot0} is from the 
Earth's velocity.

The approximations described above will serve our
purposes in exploring the features of 
the Lorentz-violating time-delay and Doppler signals.
However, 
as we discuss below in Sec.\ \ref{experimental analysis}, 
the more accurate results obtained in previous 
sections could be incorporated into a detailed
computer code for a more rigorous approach.
Furthermore, 
although we focus below on the case 
where the central body is the Sun, 
many of our results can also be applied to the 
case where the Earth or other bodies produce
the gravitational time delay and frequency shift.

\subsection{Time-delay and Doppler signals}
\label{time delay and Doppler signals}

Adopting the solar system scenario 
described above where the central
massive body is the Sun, 
we can establish the general behavior of the 
time-delay formula.
For definiteness, 
we adopt the post-newtonian 
coordinate gauge of Ref.\ \cite{qbkgrav},
setting $a_1=a_2=1$.
Though this gauge differs from the 
standard harmonic gauge at $O(3)$, 
for the $O(2)$ terms
appearing in the time-delay
expression it is equivalent.
Also, 
for times near conjunction, 
the gauge-dependent terms in \rf{td5}
will be either approximately 
constant or of order $GMb/\overline{r}$
or smaller, 
and hence negligible.

The dominant contributions
to the two-way time delay 
can be written as 
\beq
\De T_g \approx 4 GM \left[ (1+\sb^{TT}) 
\ln \left( \fr {r_e + r_p +R}{r_e +r_p -R} \right)
+\sb^{JK} \hat b^J \hat b^K \right].
\label{td6}
\eeq
To illustrate the different functional dependencies of the 
terms in Eq.\ \rf{td6} we make use of the
approximate expression in \rf{impact2}.
Up to constants, 
the expression for the time delay becomes
\bea
\De T_g &\approx& 4 GM \Big[ (1+\sb^{TT}) 
\ln \left( \fr{4 r_e r_p}{b^2_0 + {\dot b}^2_0 T^2}\right)
\nonumber\\
&&
+\sb_1 \fr {b^2_0}{b^2_0 + {\dot b}^2_0 T^2}
+ \sb_2 \fr {2 b_0 {\dot b}_0 T}{b^2_0 + {\dot b}^2_0 T^2} 
\Big],
\label{td7}
\eea
where $b_0=|\vec b_0|$ and 
${\dot b}_0=|\dot {\vec b}_0|$.
The two combinations of coefficients occurring
in Eq.\ \rf{td7} are given by
\bea
\sb_1 &=& \sb^{JK} (\hat b_0^J \hat b_0^K 
-\hat {\dot b}_0^J \hat {\dot b}_0^K),
\nonumber\\
\sb_2 &=& \sb^{JK} \hat b_0^J \hat {\dot b}_0^K,
\label{combos}
\eea
where $\hat {\dot b}_0=\dot {\vec b}_0/{\dot b}_0$.

\begin{figure}[h]
\begin{center}
\epsfig{figure=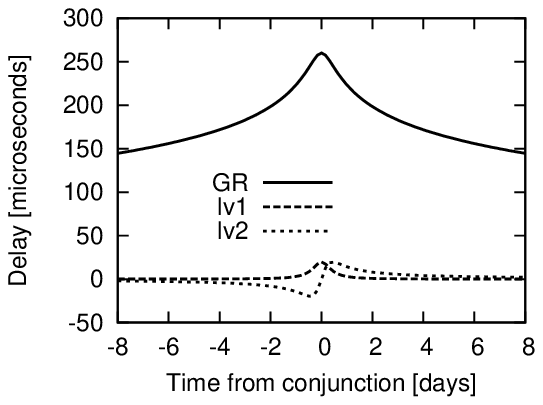,width=0.99\hsize}
\caption{\label{tdplot}
The time-delay signals occurring in Eq.\ \rf{td7} 
near solar conjunction, 
plotted with the values 
for the Cassini experiment around June 21, 2002.
The solid curve labeled GR 
gives the standard logarithmic dependence of GR, 
controlled by the combination $1+\sb^{TT}$.  
The curves labeled lv1 and lv2 
are the Lorentz-violating signals 
controlled by the combinations of 
coefficients $\sb_1$ and $\sb_2$,
respectively.}
\end{center}
\end{figure}

There are three functions that appear in expression \rf{td7}.
The first term contains the standard logarithmic
dependence present in GR, 
which is scaled by the rotational scalar
combination of coefficients 
$\sb^{TT}=\sb^{XX}+\sb^{YY}+\sb^{ZZ}$.
The second and third terms are controlled by
the anisotropic combinations of coefficients
$\sb_1$ and $\sb_2$.
To illustrate the typical behavior of the functions
occurring in \rf{td7},
we plot them in Fig.\ \ref{tdplot} for the 
case of the Cassini experiment which took place 
near the solar conjunction on June 21, 2002.
For this plot, 
we adopt the approximate values $b_0=1.6 R_{\odot}$,
${\dot b}_0 = 30 \, {\rm km/s}$,
and $GM/c^2=1.48 \, {\rm km}$, 
where $R_{\odot}$ is the Sun's radius.

The logarithmic dependence of the time-delay 
signal is well known from GR
\cite{tdold}.
The two dashed curves in Fig.\ \ref{tdplot}
represent departures from this standard behavior.
In fact, 
part of the lv2 curve controlled by the combination
of coefficients $\sb_2$
produces an advancement of the light travel time 
rather than a delay.
This may also occur with the lv2 curve
controlled by the distinct combination of 
coefficients $\sb_1$, 
if the overall sign of this combination is negative.
The peak values of the lv1 and lv2 curves 
are about $20 \mu {\rm s}$ in this example.
Note that although we are effectively setting $\sb_1=1$
and $\sb_2=1$ for the purposes of plotting, 
no specific prediction is made here.
As explained in the next subsection, 
these combinations of coefficients are expected to be much smaller
than unity given current experimental constraints. 

It is also interesting to study the 
frequency shift that corresponds to 
the time-delay signal.
For two combinations of coefficients 
for Lorentz violation in the gravitational sector, 
this signal is enhanced over the time-delay signal.
We examine the signal to $O(3)$ in the post-newtonian 
expansion and to leading order in $b/\overline{r}$, 
assuming near-conjunction times.
From Eq.\ \rf{freq2} the fractional frequency shift,
expressed in the Sun-centered frame, 
is given by
\bea
\left(\fr{\delta \nu}{\nu}\right)_g = 
\fr{8GM}{b}
[ (1+\sb^{TT}+\sb^{JK}\hat b^J \hat b^K)\dot{b}
-\sb^{JK} \hat b^J \dot{b}^K].\nonumber\\
\label{freq3}
\eea

To see some of the features of the Lorentz-violating
signals in \rf{freq3}, 
we use the approximate expression for
the impact parameter vector \rf{impact2}.
The expression for the gravitational 
fractional frequency shift becomes
\bea
\left(\fr{\delta \nu}{\nu}\right)_g & \approx & 
8GM \Big[ 
(1+\sb^{TT})\fr {{\dot b}^2_0 T}{b^2_0 + {\dot b}^2_0 T^2}
\nonumber\\
&&
+\sb_1 \fr {{\dot b}^2_0 b^2_0 T}{(b^2_0 + {\dot b}^2_0 T^2)^2}
+\sb_2 \fr {b_0 {\dot b}_0 ({\dot b}^2_0 T^2-b^2_0)}
{(b^2_0 + {\dot b}^2_0 T^2)^2}\Big].\nonumber\\
\label{freq4}
\eea

\begin{figure}[h]
\begin{center}
\epsfig{figure=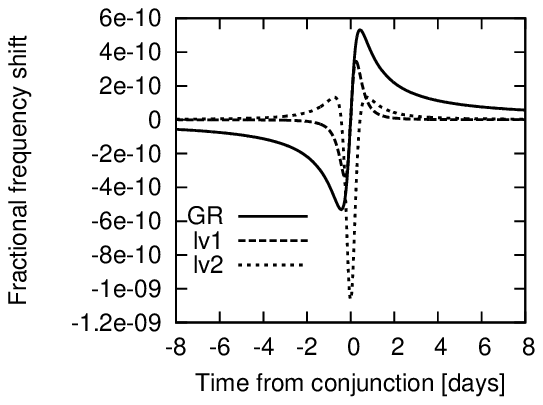,width=0.99\hsize}
\caption{\label{dplot}
The gravitational fractional frequency shift
in Eq.\ \rf{freq3} near solar conjunction, 
plotted with the values 
for the Cassini experiment around June 21, 2002.
The solid curve labeled GR 
gives the standard ${\dot b}/b$ dependence of GR, 
controlled by the combination $1+\sb^{TT}$.  
The curves labeled lv1 and lv2 
are the Lorentz-violating signals 
controlled by the combinations of 
coefficients $\sb_1$ and $\sb_2$,
respectively.}
\end{center}
\end{figure}

Just as in the time-delay case, 
three functions appear.
We plot these functions in Fig.\ \ref{dplot}, 
again using values from the Cassini experiment
and setting $\sb_1=1$ and $\sb_2=1$. 
The odd functional dependence of 
the signal controlled by the combination
$1+\sb^{TT}$ is known \cite{shapiro,bg1,bit}.
The signal controlled by $\sb_1$ resembles
the GR case, 
though its peak size is reduced.
The even functional dependence of the 
$\sb_2$ signal is qualitatively different 
from the GR case.
Note also that the maximum amplitude for this curve, 
which occurs at the conjunction time,
is about twice that of the 
peak value for the GR curve.
Also, 
as one can see qualitatively for each 
of the curves in Figs.\ \ref{dplot} and \ref{tdplot}, 
the Doppler signal is the negative of 
the time derivative of the time-delay 
signal \cite{shapiro66}.

\subsection{Experimental analysis}
\label{experimental analysis}

We discuss here key aspects of the experimental 
analysis of the time-delay and Doppler signals
for Lorentz violation in Eqs.\ \rf{td6} and \rf{freq3}.
Also, 
we make sensitivity estimates for some key 
experiments.

\begin{figure}[h]
\begin{center}
\epsfig{figure=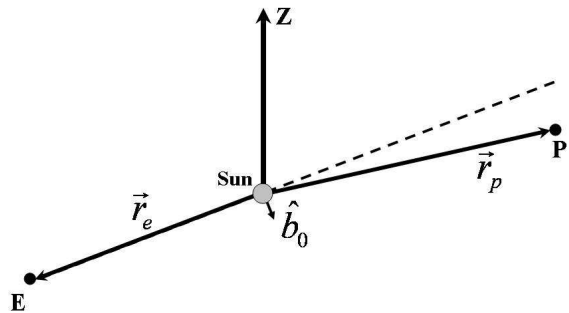,width=0.98\hsize}
\caption{\label{fig2}
Diagram illustrating the conjunction configuration 
of a typical time-delay or Doppler experiment
in the solar system.  
The Sun-centered frame $Z$ axis is shown along 
with the ecliptic plane (dashed line).}
\end{center}
\end{figure}

A ubiquitous feature of signals
for Lorentz violation is the orientation 
dependence of observable signals
\cite{light4,qbkgrav}.
Gravitational time-delay and Doppler 
tests provide no exception to this rule.
In particular,
the combinations of coefficients $\sb_1$ 
and $\sb_2$, 
controlling the lv1 
and lv2 signals in Figs.\ \ref{tdplot} and \ref{dplot},
depend on the conjunction orientation of the experiment.
To illustrate this, 
we include a sketch of the orientation 
of a typical experiment at the time of conjunction
in Fig.\ \ref{fig2}.
This figure is oriented with the Sun-centered frame $Z$
axis upwards, 
while $\vec r_e$ points in the ecliptic
to the Earth's position.
For experiments where the light signal 
comes within a few solar radii of the Sun, 
the spacecraft or planet position $\vec r_p$
is only slightly inclined to the ecliptic.

As an example of this orientation dependence, 
consider the Cassini experiment in 2002. 
Near the time of conjunction,
the Earth's velocity was pointing approximately along 
the Sun-centered frame $X$ axis (i.e., the vernal equinox
direction).
Furthermore $\hat b_0$ was pointing
very nearly perpendicular to the ecliptic.
In this case the plane of the 
illustration in Fig.\ \ref{fig2}
corresponds to the $YZ$ plane with the $X$ axis
pointing out of the page.
For this configuration we have
\bea
\hat {\dot b}_0 &\approx& (1,0,0),
\nonumber\\
\hat b_0 &\approx& (0,0.4,-0.9).
\label{cass}
\eea
This implies that the Cassini experiment 
is sensitive to the combinations of 
coefficients 
\bea
\sb_1 &\approx& 0.2 \sb^{YY}+0.8 \sb^{ZZ}-0.7\sb^{YZ}-\sb^{XX},
\nonumber\\
\sb_2 &\approx& 0.4\sb^{XY}-0.9\sb^{XZ}.
\label{cass2}
\eea

As another example, 
we suppose that the solar conjunction with the planet or spacecraft
occurs near the vernal equinox.
If this is the case, 
and the spacecraft or planet is much 
further away from the Sun than the Earth
and slightly above the $XY$ plane,
we have
\bea
\hat {\dot b}_0 &\approx& (0,-1,0),
\nonumber\\
\hat b_0 &\approx& (0,0,1).
\label{ex1}
\eea
The Sun-centered frame 
coefficients for this scenario are given by
\bea
\sb_1 &\approx& (\sb^{ZZ}-\sb^{YY}),
\nonumber\\
\sb_2 &\approx& -\sb^{YZ}.
\label{ex2}
\eea

Due to its scaling of 
the GR results
in both the time-delay and Doppler
signals,
the rotational scalar combination 
of coefficients $\sb^{TT}$ is likely 
to be constrained at the same level 
as the PPN parameter $\ga$, 
namely, 
parts in $10^5$.
However, 
care is required since 
$\sb^{TT}$ and $\ga$ are not equivalent.
In fact, 
the determination of 
the constant $GM$ may correlate with 
$\sb^{TT}$.
This is because $\sb^{TT}$ occurs at $O(2)$ 
in Newtonian gravity \cite{qbkgrav}.
For example, 
in orbital dynamics in the presence of 
$\sb^\mn$ coefficients for Lorentz
violation, 
the basic Newtonian acceleration
between two bodies is scaled by $1+5\sb^{TT}/3$.
If orbits are described as ellipses with
time-dependent orbital elements
arising from perturbations to Newtonian gravity,
the measured value $(GM)_{meas}=n^2 a^3$, 
where $n$ is the orbital frequency
and $a$ is the semimajor axis. 
Due to the presence of the $\sb^{TT}$
coefficients $(GM)_{meas}=GM (1+5\sb^{TT}/3)$.
We therefore caution the reader that care is generally
required in extracting constraints on $\sb^{TT}$.

To fit experimental data to the 
Lorentz-violating time-delay 
and Doppler signals, 
one could proceed by at least two methods.
First, 
having already fit data to a GR signal, 
one could extract constraints on SME
coefficients from time-delay or Doppler residuals.
This could be accomplished by using either 
the round-trip time-delay and Doppler formulas 
\rf{freq3} and \rf{td6} or the less accurate 
versions \rf{freq4} and \rf{td7}.
As mentioned before, 
one must bear in mind that the coefficients
for Lorentz violation $\sb^\mn$ also affect orbital
dynamics.
The effects on the orbits of the planets
due to the gravity-sector coefficients 
can be described as secular changes
over orbital time scales, 
although oscillations can also occur \cite{qbkgrav}.
However, 
these effects could in principle be avoided
with suitable filtering of the data if the focus is 
on the conjunction time scale $b/\overline{v}$.

Alternatively, 
detailed modeling of the time-delay signal
and the relevant orbital dynamics could be undertaken.
In this case, 
the one-way formula in Eq.\ \rf{td4}, 
which is valid to $O(2)$ in the 
post-newtonian expansion 
and for times far from conjunction, 
could be used appropriately for both uplink
and downlink.
The full post-newtonian equations of motion for 
the Earth and spacecraft or planet, 
and other relevant bodies  
that include the effects of 
the coefficients for Lorentz violation $\sb^\mn$
\cite{qbkgrav},
could be incorporated into the Orbital Determination
Program \cite{odp}. 
Indeed, 
data from past experiments using radar reflection from the
inner planets \cite{tdold,inner} 
could be reanalyzed to search for SME coefficients
via this second method described above.
Although many of these past experiments lack
data near conjunction, 
when the Lorentz-violating signals controlled 
by $\sb_1$ and $\sb_2$ are peaked,
they could still be useful in measuring
the rotational scalar combination $\sb^{TT}$.
Furthermore, 
detailed modeling may also reveal suppressed
dependencies of the time-delay signal 
on combinations of $\sb^\mn$ coefficients 
distinct from $\sb_1$ and $\sb_2$.

Regardless of the method adopted, 
we can make some reasonable estimates of the sensitivities
achievable in experiments.
We provide in Table \ref{estimates} estimated sensitivities
to the $3$ dominant combinations
of coefficients in the time-delay and Doppler
experiments for some past and future experiments.
We include the Cassini experiment 
and some key future tests. 
The estimates are order of magnitude only and are based on 
the peak values of the Lorentz-violating signals
discussed above and the approximate accuracy of each 
experiment referenced, 
when available. 
For example, 
the peak value of the $\sb_2$ Doppler 
signal for the Cassini experiment is about $10^{-9}$, 
while the Allan deviation for this experiment
is about $10^{-14}$ \cite{bit}, 
indicating a sensitivity of parts in $10^5$.
However, 
data from the time period when the $\sb_2$ signal peaked 
in the 2002 conjunction ($T \approx 0$ in Fig.\ \ref{dplot})
is not available, 
so the sensitivity to $\sb_2$ is more likely to be parts in $10^4$. 
On the other hand, 
it appears likely that a suitable fitting of Cassini data
could place the first constraints on the rotational
scalar combination $\sb^{TT}$ at the $10^{-5}$ level.
For the time-delay signals, 
the sensitivity to the $\sb_1$ and $\sb_2$ coefficients
is reduced by about a factor of $10$ or more, 
as indicated in Fig.\ \ref{tdplot},
and this reduction in sensitivity
is included in Table \ref{estimates}.

\begin{table}
\begin{center}
\begin{tabular}{|l|c|c|c|c|}
\hline
Experiment & $\sb^{TT}$  & $\sb_1$
& $\sb_2$ & Ref. \\
\hline
\multicolumn{5}{|c|}{Time-delay signal}\\
\hline
Cassini & $10^{-5}$ & $10^{-3}$ 
& $10^{-4}$ & \cite{bit} \\
\hline
Odyssey & $10^{-7}$ & $10^{-6}*$ 
& $10^{-6}*$ & \cite{odyssey} \\
\hline
ASTROD & $10^{-8}$ & $10^{-7}*$ 
& $10^{-7}*$ & \cite{astrod} \\
\hline
BEACON & $10^{-9}$ & $10^{-8}*$ 
& $10^{-8}*$ & \cite{beacon} \\
\hline
\multicolumn{5}{|c|}{Doppler signal}\\
\hline
Cassini & $10^{-5}$ & $10^{-4}$ & $10^{-4}$
& \cite{bit} \\
\hline
Odyssey & $10^{-7}$ & $10^{-7}*$ 
& $10^{-7}*$ & \cite{odyssey} \\
\hline
\end{tabular}
\caption{\label{estimates}
Crude estimates of attainable sensitivities
in some key experiments for the time-delay 
and Doppler signals.}
\end{center}
\end{table}

Proposals have been put forth for future experiments
that measure to impressive accuracies
the time delay from the Sun and even the Earth. 
We have included sensitivity estimates for the Odyssey, 
ASTROD, 
and BEACON experiments in Table \ref{estimates}.
Although in some cases it may be difficult to directly
measure the fractional frequency shift \cite{b07}, 
nonetheless we include some estimates in the table 
because of the possibility of increased sensitivity
to SME coefficients from the Doppler signal over the
time-delay signal.
The experiments in Table \ref{estimates} 
are by no means an exhaustive list.
Also of possible interest are proposals for 
measuring the light-bending effect 
such as SIM \cite{sims} and LATOR \cite{lator}, 
other proposed experiments \cite{bepicolombo,gaia},
as well as existing accumulated data from Earth satellites
\cite{gps}. 
Though it lies beyond the scope of the present work, 
it would also be of interest to obtain the corresponding
light-bending signal controlled by the $\sb^\mn$ 
coefficients.

Note that current constraints
on the off diagonal components $\sb^{XY}$, 
$\sb^{YZ}$, 
and $\sb^{XZ}$ are at the level of $10^{-8}$ from 
atom interferometry \cite{gravi}.
Two combinations of these and other $\sb^{JK}$ 
coefficients are also constrained by lunar laser ranging
at the $10^{-10}$ level \cite{llr}.
Thus, 
if future experiments can measure the 
peak behavior of the $\sb_2$ set of coefficients
in the Doppler signal to better than parts in $10^8$, 
they may produce measurements of coefficients 
competitive with or better than previous experiments.
The $*$ label next to the 
estimated sensitivities in Table \ref{estimates}
indicates the requirement of measuring the 
peak behavior of the time-delay and Doppler signals.
Finally we note that the $\sb^{TT}$ coefficient does not appear
at leading order in laboratory and 
orbital tests \cite{qbkgrav} 
and so time-delay and Doppler tests are likely 
to be among the most sensitive to this coefficient.

\section{Summary}
\label{summary}

In this work,
we have analyzed Lorentz-violating 
corrections to the gravitational time-delay and Doppler
signals in General Relativity,
in the context of the gravitational 
sector of the minimal SME.
We established general integral formulas
for the deviation of a light ray from 
a straight line path that are valid
in the linearized gravity limit.
Our main results are analytical formulas
for the light travel time and frequency shift
for a light signal sent between two observers
past a massive central body in the presence 
of gravity sector coefficients $\sb^\mn$.
We obtained the one-way results in 
Eqs.\ \rf{td4} and \rf{fr6} and 
the round-trip signals in 
Eqs.\ \rf{td5} and \rf{freq2}.

The Lorentz-violating signals were studied 
for solar system experiments involving light signals
sent between the Earth and a planet or spacecraft near 
solar conjunction. 
It was determined that the dominant signals are controlled
by the combinations of 
coefficients $1+\sb^{TT}$, 
$\sb_1$, 
and $\sb_2$.
In terms of Sun-centered frame coefficients, 
the combinations $\sb_1$ and $\sb_2$ will 
vary for different experiments.
We obtained sensitivity estimates
for key existing and future experiments
which are summarized in Table \ref{estimates}.
Time-delay and Doppler experiments
could prove crucial in measuring the 
elusive scalar coefficient $\sb^{TT}$,
to better than parts in $10^5$. 
Future highly sensitive time-delay and 
Doppler tests may be able to measure other 
coefficients in the subset $\sb^{JK}$ 
with sensitivities competitive with other 
existing experiments.


\begin{thebibliography}{}
\bibitem{scf}
A detailed figure depicting the Sun-centered frame
can be found in Fig.\ $2$ of Ref.\ \cite{qbkgrav}.
\end{thebibliography}
\end{document}